\documentclass[aps,prl,floatfix,twocolumn,tightenlines,superscriptaddress,amsmath,amssymb]{revtex4-1}

\usepackage{graphicx}
\usepackage{amssymb}
\usepackage{color}
\usepackage{psfrag}
\usepackage{ifsym}

\newcommand{\bra}[1] {\langle #1 |}
\newcommand{\ket}[1] {| #1 \rangle}

\begin{document}
\title{Hardy's paradox and violation of a state-independent Bell inequality in time}

\author{Alessandro Fedrizzi}
\affiliation{Centre for Engineered Quantum Systems and Centre for Quantum Computer and Communication Technology, School of Mathematics and Physics, University of Queensland, 4072 Brisbane, QLD, Australia}
\author{Marcelo P. Almeida}
\affiliation{Centre for Engineered Quantum Systems and Centre for Quantum Computer and Communication Technology, School of Mathematics and Physics, University of Queensland, 4072 Brisbane, QLD, Australia}
\author{Matthew A. Broome}
\affiliation{Centre for Engineered Quantum Systems and Centre for Quantum Computer and Communication Technology, School of Mathematics and Physics, University of Queensland, 4072 Brisbane, QLD, Australia}
\author{Andrew G. White}
\affiliation{Centre for Engineered Quantum Systems and Centre for Quantum Computer and Communication Technology, School of Mathematics and Physics, University of Queensland, 4072 Brisbane, QLD, Australia}
\author{Marco Barbieri}
\affiliation{Groupe d'Optique Quantique, Laboratoire Charles Fabry, Institut d'Optique, \\
CNRS, Universit\'e Paris-Sud XI, Campus Polytechnique, RD 128, 91127 Palaiseau cedex, France}

\begin{abstract}
Tests such as Bell's inequality and Hardy's paradox show that joint probabilities and correlations between distant particles in quantum mechanics are inconsistent with local realistic theories. Here we experimentally demonstrate these concepts in the time domain, using a photonic entangling gate to perform nondestructive measurements on a single photon at different times. We show that Hardy's paradox is much stronger in time and demonstrate the violation of a temporal Bell inequality independent of the quantum state, including for fully mixed states.
\end{abstract}
\maketitle

Quantum mechanics depicts a world with fuzzier contours than our intuitive mind would suggest. In our common experience, we would naively picture a measurement as a way of revealing some objective properties. 
This view is disproved by several counterexamples, of which the most common is provided by an entangled system: the correlations between measurement outcomes can not  be explained by a theory assuming that each subsystem has values determined independently of a measurement itself. Well-proven tests such as Hardy's paradox \cite{hardy1993nft} and Bell's inequality \cite{bell1966phv,aspect2007qmt} capture these features of spatial entanglement. 

This inconsistency can be expressed in a different setting; as pointed out by Legget and Garg in their seminal paper \cite{leggett1985qmm}, one can consider correlations between measurements on the \emph{same object} occurring at \emph{different times}. Their aim was to find a particular instance where a realistic view was untenable, which has subsequently been the subject of numerous theoretical \cite{jordan2006lgi,williams2008wvl,kofler2008cqv,barbieri2009mml} investigations and experimental demonstrations \cite{goggin2009vlg,xu2009eit,palacios2010evb,dressel2011evt,waldherr2011vtb}. In a more general context, temporal quantum phenomena, in particular ``entanglement in time'', have subsequently been studied in \cite{brukner2004qet, lapiedra2006jrb, fritz2010qct}. 

Here, we report the first experimental investigations of these concepts: that Hardy's paradox is much stronger in time \cite{fritz2010qct}, and that a temporal Bell inequality can be state independent---it can be violated by all quantum states, even fully mixed ones \cite{brukner2004qet}. Our experiment highlights surprising aspects of quantum foundations---such as all quantum states are entangled in time. Furthermore, entanglement in time might inspire new protocols in quantum information, communication and control \cite{brukner2004qet}.

 \begin{figure}[t!]
  \begin{center}
 \includegraphics[width=\columnwidth]{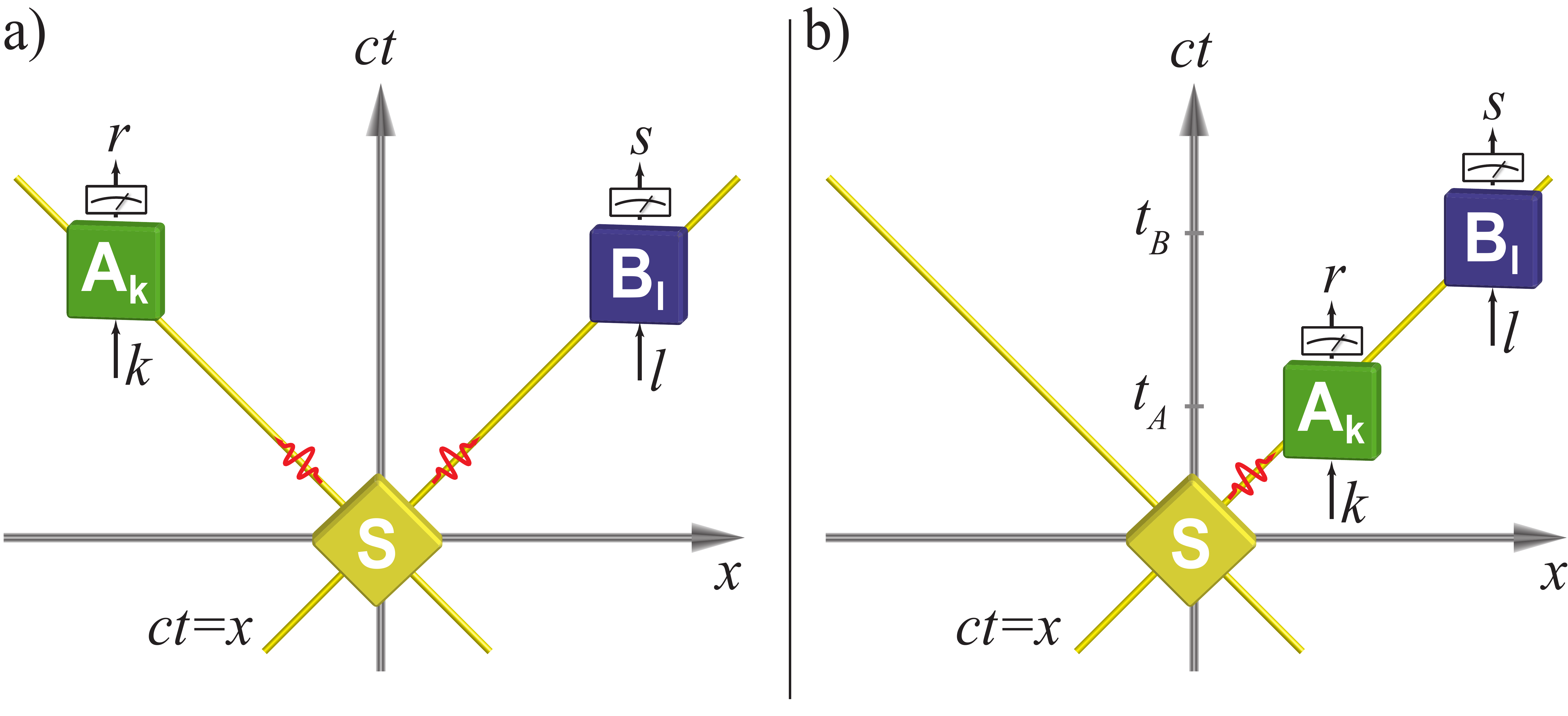}
  \end{center}
\caption{Thought experiment for the violation of local realistic theories. (a) Spatial scenario: A source $S$ emits two (entangled) qubits, which are sent to two remote observers A and B. Each subsystem is subject to two measurements $A_{k}$ and $B_{l}$, where $k$ and $l$ denote the measurement settings at different sites. The outcomes of individual measurements are labelled $r$ and $s$. (b) Temporal scenario for the violation of non-invasive, realistic theories. A single system is subjected to two measurements $A_{k}$ and $B_{l}$ , in this case occurring  at different times $t_{B}>t_{A}$.} 
  \label{fig:spacetime}
\end{figure}

Consider a quantum system located at two points in spacetime, $A$ and $B$, where a quantum particle exists at each point. Our classical view of such a system is based on two assumptions: (i) realism, that the particle at each point has definite properties prior to, and independent of measurements; and (ii) non-disturbance, that results of measurements at $A$ are not influenced by measurements at $B$, and vice-versa. In the spatial case, Fig.~\ref{fig:spacetime} (a), there are separate particles at $A$ and $B$, and special relativity ensures that disturbances cannot propagate between them faster than the speed of light. Thus tenet (ii)---now termed \emph{locality}---can be enforced by a space-like separation. In the temporal case, Fig.~\ref{fig:spacetime} (b), a single particle is measured at different times, $t_{A}$ and $t_{B}$. Because these measurements lie within each other's light cone, locality in the traditional sense cannot be enforced. We can however still define a classical picture in this scenario: It is reasonable to assume that, while signalling cannot be avoided, a measurement can be performed such that it does not influence the outcome of a measurement on the same system at a later (or earlier) time. This hypothesis of \emph{measurement noninvasiveness} was originally introduced by Leggett and Garg \cite{leggett1985qmm}, motivated by their interest in macroscopic systems. Its applicability to microscopic objects has been discussed \cite{goggin2009vlg}. 

Replacing locality in time with noninvasiveness might appear controversial when compared to an invasive theory such as quantum mechanics. An alternative is to consider perfect repeatability of a quantum measurement, which is compatible with quantum mechanics and allows the construction of hidden variable models identical to those originating from the assumption of noninvasiveness \cite{lapiedra2006jrb,fritz2010qct}.
%

Despite the fact that two-body correlations in space and time are mathematically equivalent \cite{brukner2004qet}, there are remarkable differences between measurements on quantum systems in the two domains. The first can be found in the temporal version \cite{fritz2010qct} of Hardy's paradox \cite{hardy1993nft,boschi1997lpn,white1999nes,barbieri2005ttn,irvine2005rht,lundeen2009ejw,yokota2009doh}. It
describes a paradoxical situation in which quantum mechanics allows a set of probabilities which are logically inconsistent within a classical framework. Consider two observers, Alice and Bob, sharing a single system on which they conduct a joint sequential measurement with the choices $A_{k}$ and $B_{l}$, with $k,l=\{0,1\}$, at two different times, Fig.~\ref{fig:spacetime} (b). The measurements are dichotomic, with the possible outcomes $r,s=\{0,1\}$. The probability of a result $r$ for Alice and $s$ Bob is $P(r,s|l,k)$.  

Now consider the following set of outcome probabilities for different measurement choices on this system:
\begin{align}
\label{eq1}P(1,1|1,1)>0,\\
\label{eq2}P(1,0|1,0)=0,\\
\label{eq3}P(0,1|0,1)=0, \\
\label{eq4}P(1,1|0,0)=0.
\end{align}
Equation~\eqref{eq1} predicts the existence of events that give the outcome $r=1$, $s=1$ for a joint measurement $A_{1}$, $B_{1}$. For a system obeying realism and nondisturbance, these values of $r$ and $s$ are defined before the measurement, and the choice of operator $A_k$  cannot possibly affect the outcome of $B_l$, and vice versa. Thus, due to \eqref{eq2}, had we instead chosen $B_0$, we would certainly have found a result $s=1$. In the same way, according to Eq.~\eqref{eq3}, we would have observed $r=1$ for the alternative choice $A_0$. This however demands the occasional occurrence of events with the outcome $r=1$, $s=1$ for choices $A_{0}$ and $B_{0}$, which is clearly inconsistent with \eqref{eq4}~\cite{hardy1993nft, fritz2010qct}.


\begin{figure}[t]
  \begin{center}
 \includegraphics[width=\columnwidth]{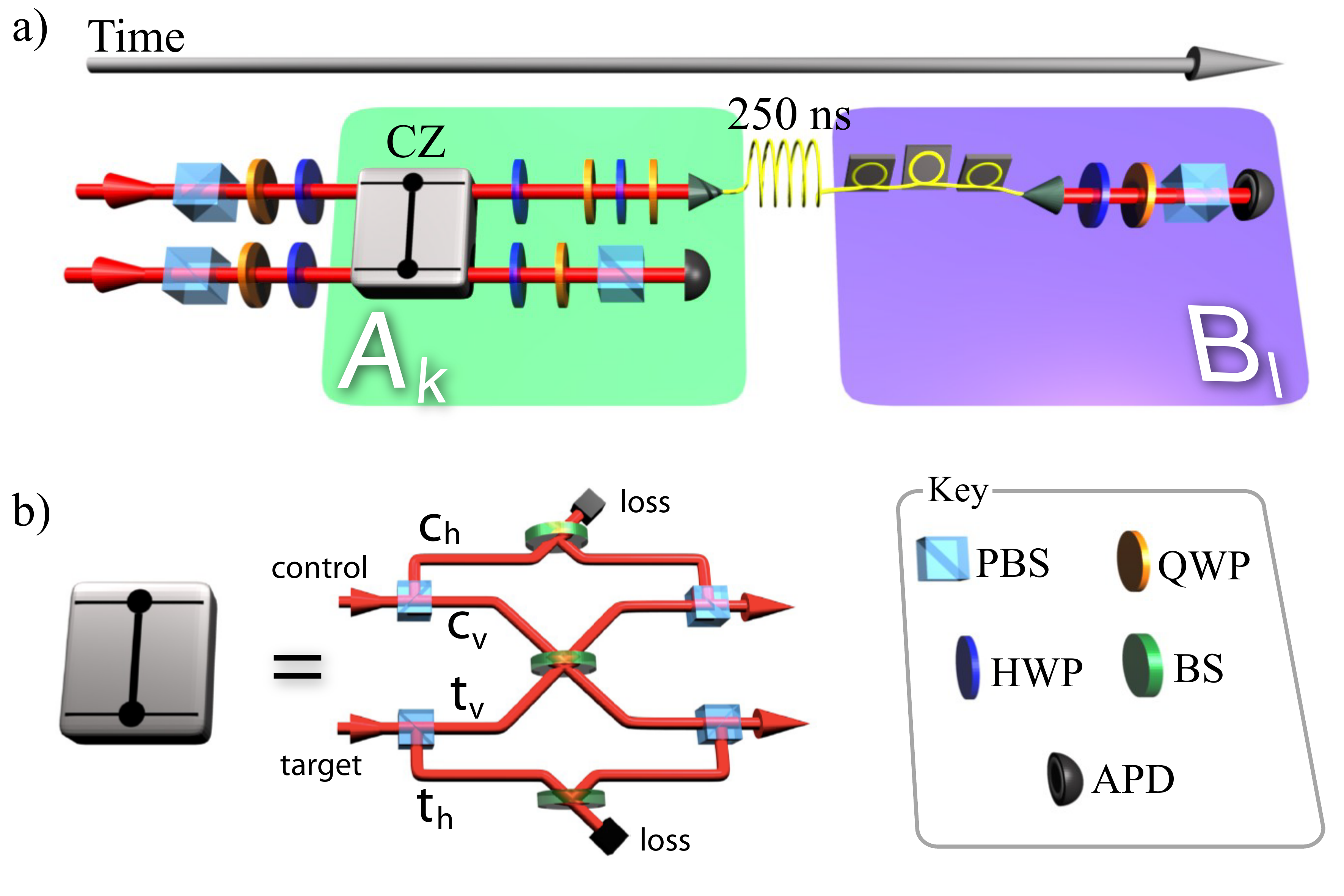}
  \end{center}
\caption{Experimental scheme. a) Temporal measurements. The signal and meter qubit are encoded in orthogonal polarization states of two single photons, which are created via spontaneous parametric down-conversion in a nonlinear crystal, pumped by a pulsed (76 MHz, 200 fs), frequency-doubled Ti:Sapphire laser at $\lambda{=}820$~nm. States are prepared with polarising beamsplitters (PBS), a quarter- (QWP) and a half-wave plate (HWP). The signal photon passes a controlled-phase gate (\textsc{cz}), where it acts as the control qubit, with the meter photon being the target. Behind the gate, we analyze the meter photon polarization and detect it with a single-photon avalanche photo diode (APD), implementing the first measurement $A_{k}$. Two HWPs (one incorporated into the preparation stage) set the basis for this non-destructive measurement. The signal is stored in a $50$ meter long fiber spool and, after $A_{k}$ is concluded, measured projectively, implementing $B_{l}$. A fiber polarization controller and a combination of wave plates compensate for polarization rotation in the fiber. A coincidence logic analyzes detection events within a time window of $4.4$~ns. b) The \textsc{cz} gate in detail, here shown in dual-rail representation. We realize it with a single partially polarising beam splitter (PPBS), with transmittivities $\eta_{H}=1/3$ ($\eta_{V}=1$) for the H (V) polarisation \cite{langford2005dse,kiesel2005loc,okamoto2005doq}. Quantum interference results in a relative $\pi$ phase shift of the vertical polarization components $\ket{V}_{s}\ket{V}_{m}$. The correct functioning is heralded by a coincidence count between the two output arms of the PPBS, which occurs with probability $1/9$.} 
  \label{fig:setup}
\end{figure}

Quantum mechanics, of course, resolves the paradox \cite{fritz2010qct}.
Consider a single two-level quantum system (qubit) prepared in the state $| 0 \rangle$. With the Pauli measurements $A_{0}{=}B_{1}{=}-Z$, and $A_{1}{=}B_{0}{=}X$, where $Z$ and $X$ are the Pauli operators corresponding to the measurement along the $z$ and, respectively, $x$ directions on the Bloch sphere, it satisfies the equations (\ref{eq1})-(\ref{eq4}), with $P(1,1|1,1){=}0.25$. 

In principle, a single observation of a detection event for the settings $k,l=1$ (\ref{eq1}) would---in the absence of detections for settings (\ref{eq2})---provide a compelling proof that nature does not obey the classical worldview established by the assumptions of realism and noninvasiveness \cite{lapiedra2006jrb,fritz2010qct}. However, even in an ideal scenario, zero probabilities can only ever be established to within an error governed by the number of measurement runs. In practice, we have to deal with imperfect states, measurements and detectors, which exacerbates this problem. We can instead, following Mermin \cite{mermin1994qmr}, place a bound on $P(1,1|1,1)$, given the measured values of the other probabilities: 
\begin{align}
\label{eq:inequality}
\mathcal{H}=&P(1,1|1,1)-P(1,1|0,0)\\
-&P(1,0|1,0)-P(0,1|0,1)\leq0. \nonumber
\end{align}

We test this inequality in a two-photon experiment, see Fig.~\ref{fig:setup}a). A system qubit is encoded in the polarisation of a single photon; horizontal (H) and vertical (V) polarisations determine the $z$-axis of the Bloch sphere. We implement the first, necessarily non-destructive, measurement using a non-deterministic, photonic controlled-phase (\textsc{cz}) gate, Fig.~\ref{fig:setup}b). It acts on two polarisation qubits, the signal $\ket{\psi}_{s}$, and the meter $\ket{\phi}_{m}$. The state of the signal qubit controls the meter, acting as the target qubit. An input state $\ket{V}_{s}\ket{D}_{m}$, for example, undergoes the controlled rotation $\ket{V}_{s}\ket{D}_{m}\rightarrow \ket{V}_{s}\ket{A}_{m}$ while $\ket{H}_{s}\ket{D}_{m}\rightarrow \ket{H}_{s}\ket{D}_{m}$ \cite{definition}. The polarisation of the signal can then be inferred by its action on the meter \cite{goggin2009vlg,pryde2004mpq,barbieri2009cvs}. If the arbitrary state $\ket{\psi}_{s}\ket{D}_{m}$ is injected, we can measure $Z$ on the signal just by observing whether the meter has been rotated or not. Arbitrary measurements can be chosen by rotating the signal before the gate. This rotation must be undone at the gate output, as shown in Fig.~\ref{fig:setup}.

We experimentally measured $P(1,1|1,1)=0.2372\pm0.0040$, $P(1,1|0,0)=0.0181\pm0.0008$, $P(1,0|1,0)=0.0190\pm0.0013$, $P(0,1|0,1)=0.0070\pm0.0005$, yielding 
\begin{equation}
\mathcal{H}_\textrm{exp}{=}0.193{\pm}0.004, \nonumber
\end{equation}
which violates inequality~\eqref{eq:inequality} by $45$ standard deviations.

The key feature is that this temporal version of Hardy's proof is considerably stronger than its spatial analogue, where the left-hand side of \eqref{eq:inequality} can be no greater than $\sim0.09$ \cite{mermin1994qmr}; our results surpass this limit by more than $24$ standard deviations. The violation of Hardy's inequality in time can be achieved by any pure quantum state, provided that the observables are chosen appropriately. 

Surprisingly, and in stark contrast to its spatial analogue, such pure states are not required for the Clauser-Horne-Shimony-Holt (CHSH) form of a temporal Bell inequality \cite{clauser1969pet}. Unlike Hardy's paradox, the CHSH inequality considers correlations between points $A$ and $B$. The two results will be correlated whenever ${r=s}$ and anti-correlated in the other case. Hence, the correlation function for two observables $A_{k}$ and $B_{l}$ is 
\begin{equation}
C_{k,l}{=}\sum_{r,s}(-1)^{r+s}P(r,s|k,l). 
\end{equation}
By invoking realism and noninvasiveness to establish a bound on correlations one can then define the temporal Bell inequality \cite{brukner2004qet}:
\begin{equation}
\label{eq:chsh}
S=|C_{0,0}+C_{1,0}+C_{0,1}-C_{1,1}|\leq2,
\end{equation}
which has the same form as the CHSH inequality in the spatial domain \cite{clauser1969pet}.

For a quantum state $\rho$, the expectation value of $C_{k,l}$ is given by
\begin{equation}
\label{eq:corr}
C_{k,l}=\mathrm{Tr}(\rho\cdot\frac{1}{2}[A_{k},B_{l}]_{+}),
\end{equation}
where $[A_{k},B_{l}]_{+}$ is the anti-commutator of the measurement operators \cite{fritz2010qct}. For a single qubit, a maximal violation of inequality \ref{eq:chsh}, $S_{QM}=2\sqrt{2}$, can be obtained by choosing appropriate measurements on the Bloch sphere. We select the same operators as in spatial CHSH experiments: $A_{0}{=}Z$, $A_{1}{=}X$, $B_{0}{=}(Z{+}X)/\sqrt{2}$ and $B_{1}{=}(Z{-}X)/\sqrt{2}$. Remarkably, the correlators $C_{k,l}$, Eq.~\eqref{eq:corr}, and thus the parameter $S$ do not depend on the choice of the quantum state $\rho$, but only on the measurement operators. If we denote $\vec a_{k}$ and $\vec b_{l}$ the directions associated with $A_{k}$ and $B_{l}$, the correlation is simply given by \cite{brukner2004qet} $C_{k,l}{=}\vec a_{k}\cdot \vec b_{l}$. Note that this is not the case for the Leggett-Garg form of a temporal Bell inequality \cite{leggett1985qmm,jordan2006lgi,williams2008wvl} which has recently been tested experimentally \cite{goggin2009vlg,xu2009eit,palacios2010evb}.

\begin{figure}[t]
  \begin{center}
 \includegraphics[width=.88\columnwidth]{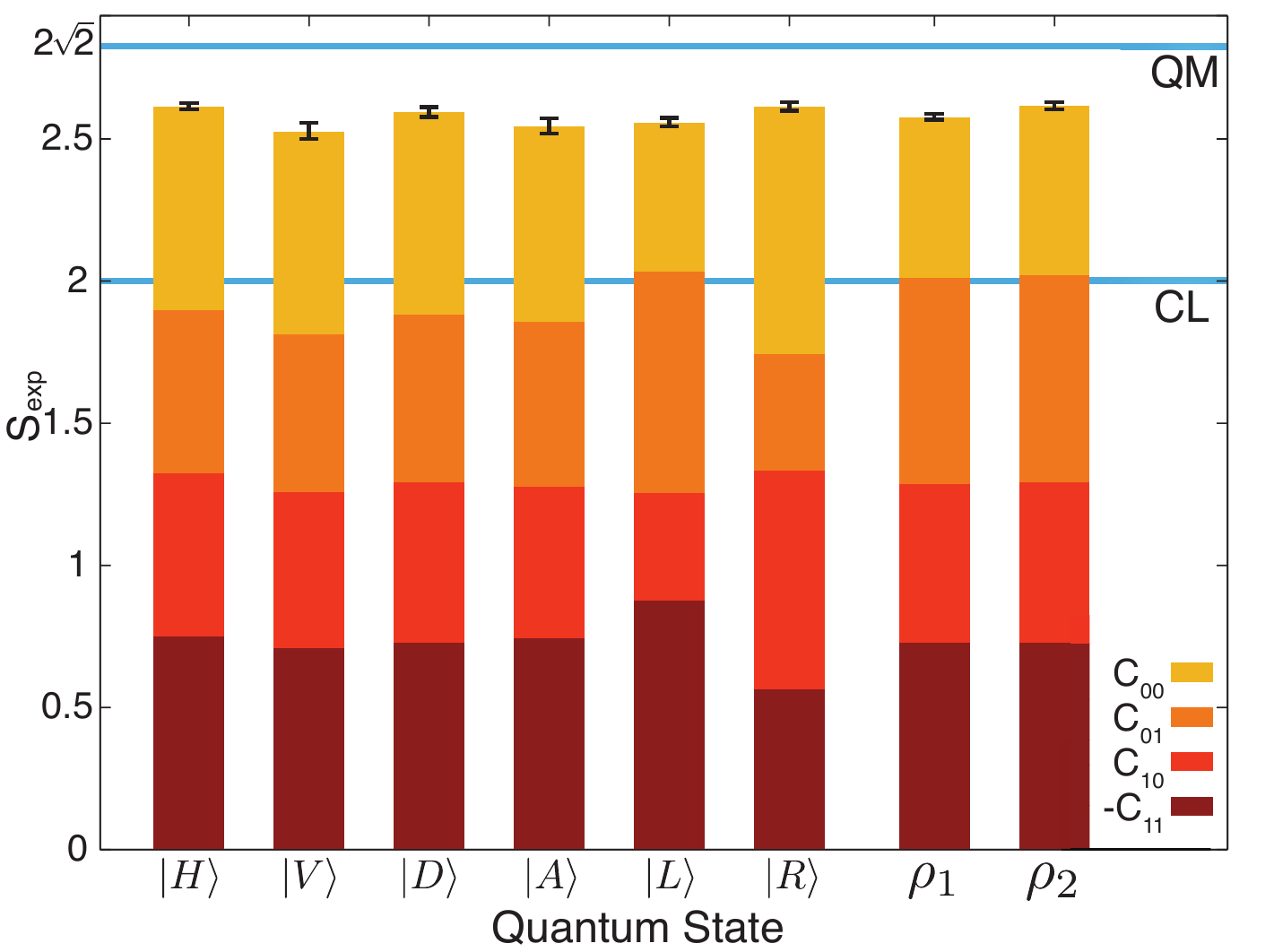}
  \end{center}
\caption{Experimental violation of the state-independent temporal Bell inequality (\ref{eq:chsh}). The classical limit is indicated by CL and the maximal achievable quantum value by QM. The first six bars correspond to pure signal states the remaining two to mixed inputs, as explained in the main text. The latter were obtained by switching the signal state between states $| D \rangle$ and $| A \rangle$ while measurements were performed. The relative integration for $| D \rangle$ and $| A \rangle$ were chosen according to the target purity of $0.5$ for $\rho_{1}$ and $0.75$ for $\rho_{2}$. The mixed states were verified via single-qubit tomography.} 
  \label{fig:results}
\end{figure}

The experimental results are summarised in Fig. \ref{fig:results}. We tested the temporal Bell inequality, Eq.~\eqref{eq:chsh}, for a total of eight states; six (almost) pure input states, $\{\ket{H}$, $\ket{V}$, $\ket{D}$, $\ket{A}$, $\ket{L}$, $\ket{R}\}$; one mixed state $\rho_{1}{\sim}(0.84\ket{H}\bra{H}{+}0.16\ket{V}\bra{V})$ with purity $\mathcal{P}{=}0.74{\pm}0.01$, and one fully mixed state $\rho_{2}{\sim}1/2(\ket{H}\bra{H}{+}\ket{V}\bra{V})$ with purity $\mathcal{P}{=}0.50{\pm}0.01$. The experimentally obtained S-parameter for these states was, on average, 
\begin{equation}
S_\textrm{exp}=2.58\pm0.03, \nonumber
\end{equation}
which violates inequality (\ref{eq:chsh}) by $19$ standard deviations. It is quite remarkable that we get a clear violation even with a fully mixed state, for which one would---intuitively---not expect any evident quantum signature.

The observed Bell value corresponds to a two-point visibility of $0.91\pm0.01$. The less-than-maximal violation of the temporal inequality is due to imperfect measurement, which is mainly limited by less-than-ideal two-photon interference in the gate. We can assess the measurement performance by performing quantum process tomography \cite{obrien2004qpt} on our gate. The experimental process $\chi_\textrm{exp}$ associated with the measurement has a purity of $92.4\pm0.2\%$ and a fidelity with an ideal $\textsc{cz}$ process of $93.7\pm0.1\%$. The error bounds are calculated from $10$ Monte Carlo runs assuming Poissonian photon count statistics. The upper bound on the CHSH value (\ref{eq:chsh}), calculated from  $\chi_\textrm{exp}$, is $2.54\pm0.01$ averaged over all input states and, within error, in good agreement with the measured value. For the Hardy inequality (\ref{eq:inequality}), the estimated bound is $0.184\pm0.003$---slightly below the respective experimental result, which is most likely due  to temporal drift in the optical setup.

The study of temporal quantum phenomena offers a new perspective for quantum information. The authors of \cite{brukner2004qet}, e.g., propose a temporal quantum communication complexity protocol where temporal entanglement provides a memory advantage over classical information. It is conceivable that we can also find classically-impossible, or more efficient quantum communication tasks based on the strong quantum signature of temporal probabilities. 

Our investigation also raises more fundamental questions. The first concerns the potential link between temporal quantum phenomena and contextuality \cite{kochen1967phv}, another example of a state-independent incosistency of the classical and quantum world \cite{kirchmair2009sie}. Future efforts will investigate possible connections between contextuality and invasiveness.

A different question is if, and to which degree, non-invasiveness could be relaxed while still allowing violation by quantum mechanics. Intriguingly, for the Leggett-Garg inequality, the connection between the measurement strength and the amount of violation is not straightforward: the less invasive the measurement, the higher the violation \cite{williams2008wvl,goggin2009vlg}. Ultimately, the fundamental differences of quantum effects in the two domains may teach us more about the structure of space and time and the abstract formalism of quantum theory \cite{brukner2004qet}.

 \begin{acknowledgments}
We thank T. Paterek, C. Branciard and G. Milburn for insightful discussions. We acknowledge financial support from the ARC Centre of Excellence, Discovery and Federation Fellow programs and an IARPA-funded US Army Research Office contract. M.B. is supported by the Marie Curie contract PIEF-GA-2009-236345-PROMETEO of the E.U. Commission
\end{acknowledgments}

\end{document}